\documentclass[aps,reprint,twocolumns,pre,superscriptaddress,floatfix,notitlepage,nofootinbib]{revtex4-1}
\usepackage{amssymb,amsmath,amsfonts}
\usepackage{mathtools}
\usepackage{physics}
\usepackage{array}
\usepackage{graphicx,epsfig,xcolor}
\usepackage[colorlinks=true,citecolor=blue,linkcolor=red]{hyperref}
\usepackage{newtxtext,newtxmath}
\begin{document}

\newcommand{\hatmath}[1]{\hat{\mathcal{#1}}}

\title{Exceptional Spectral Phase  in a Dissipative Collective Spin Model}

\author{\'{A}lvaro Rubio-Garc\'{i}a}
  \email[]{alvaro.rubio@csic.es}
  \affiliation{Instituto de F\'{\i}sica Fundamental, IFF-CSIC, Serrano 113, 28006 Madrid, Spain}

\author{\'{A}ngel L. Corps}
  \email[]{angel.l.corps@csic.es}
  \affiliation{Instituto de Estructura de la Materia, IEM-CSIC, Serrano 123, 28006 Madrid, Spain}
  \affiliation{Grupo Interdisciplinar de Sistemas Complejos (GISC), Universidad Complutense de Madrid, Av. Complutense s/n, E-28040 Madrid, Spain}

\author{Armando Rela\~no}
  \affiliation{Grupo Interdisciplinar de Sistemas Complejos (GISC), Universidad Complutense de Madrid, Av. Complutense s/n, E-28040 Madrid, Spain}
  \affiliation{Departamento de Estructura de la Materia, F\'{i}sica T\'{e}rmica y Electr\'{o}nica, Universidad Complutense de Madrid, Av. Complutense s/n, E-28040 Madrid, Spain}

\author{Rafael A. Molina}
  \affiliation{Instituto de Estructura de la Materia, IEM-CSIC, Serrano 123, 28006 Madrid, Spain}

  \author{Francisco P\'{e}rez-Bernal}
\affiliation{Departamento de Ciencias Integradas, Facultad de Ciencias Experimentales, Centro de Estudios Avanzados en F\'{i}sica, Matem\'{a}ticas y Computaci\'{o}n, Unidad Asociada GIFMAN, CSIC-UHU, Universidad de Huelva, 21071 Huelva, Spain}
\affiliation{Instituto Carlos I de F\'{i}sica Te\'{o}rica y Computacional, Universidad de Granada, 18071 Granada, Spain}

\author{Jos\'{e}-Enrique Garc\'{i}a-Ramos}
\affiliation{Departamento de Ciencias Integradas, Facultad de Ciencias Experimentales, Centro de Estudios Avanzados en F\'{i}sica, Matem\'{a}ticas y Computaci\'{o}n, Unidad Asociada GIFMAN, CSIC-UHU, Universidad de Huelva, 21071 Huelva, Spain}
\affiliation{Instituto Carlos I de F\'{i}sica Te\'{o}rica y Computacional, Universidad de Granada, 18071 Granada, Spain}

\author{Jorge Dukelsky}
  \email[]{j.dukelsky@csic.es}
  \affiliation{Instituto de Estructura de la Materia, IEM-CSIC, Serrano 123, 28006 Madrid, Spain}

\date{\today}

\begin{abstract}

We study a model of a quantum collective spin weakly coupled to a spin-polarized Markovian environment and find that the spectrum is divided into two regions that we name normal and exceptional Liouvillian spectral phases. In the thermodynamic limit, the exceptional spectral phase displays the unique property of being made up exclusively of second order exceptional points. As a consequence, the evolution of any initial density matrix populating this region is slowed down and cannot be described by a linear combination of exponential decays. This phase is separated from the normal one by a critical line in which the density of Liouvillian eigenvalues diverges, a phenomenon analogous to that of excited-state quantum phase transitions observed in some closed quantum systems. In the limit of no bath polarization, this criticality is transferred onto the steady state, implying a dissipative quantum phase transition and the formation of a boundary time crystal.

\end{abstract}

\maketitle

{\em Introduction.-} Real quantum systems are always in contact with environments that induce dissipation and/or decoherence. There is a growing interest in understanding these effects, which, if controlled, can be used as a resource for new devices and applications \cite{Diehl2008,San-Jose2016,Liu2020}. For this purpose, the Gorini-Kossakowski-Sudarshan-Lindblad (GKSL) approximation for Markovian environments becomes especially relevant. It is given by the Lindblad master equation
\begin{equation}
  \mathcal{L}(\rho) := -i\left[\hat{H},\rho\right]+\sum_{i}\left( \hat{L}_i\rho \hat{L}_i^{\dagger}-\frac{1}{2}\left\{\hat{L}_i^{\dagger}\hat{L}_i, \rho\right\}\right) = \frac{\partial \rho}{\partial t},
  \label{eq:lindblad}
\end{equation}
where $\hat{H}$ is the Hamiltonian describing the unitary evolution of the closed system, and $\hat{L}_i$ are the Lindblad jump operators \cite{Linblad1976,Gorini1976}. These operators define the interaction between the system and the environment. The evolution of the density matrix, $\rho$, is determined by the spectrum of the Liouvillian superoperator, $\mathcal{L}$, which always has at least one eigenstate with a zero eigenvalue, $\mathcal{L}(\rho_{SS})=0$, determining the steady state (SS).

As the Liouvillian is non-Hermitian, it allows for the existence of exceptional points (EPs) whereby two or more eigenvalues and eigenvectors of $\mathcal{L}$ coalesce \cite{Heiss2012,Ueda2020}. While Hamiltonian EPs, produced by non-Hermitian Hamiltonians, have been extensively  studied, this is not the case for Liouvillian EPs. They have been only discussed in a few cases \cite{Minganti2019,Hatano2019,Naghiloo2019, Zambrini2020,Khandelwal2021,Claeys2022}.

In this Letter we define and characterize what we call an Exceptional Spectral Phase (ESP). It consists in a region of the Liouvillian spectrum where all eigenvectors coalesce in pairs in the thermodynamic limit (TL). The line demarcating the boundary between the ESP and the normal spectral phase displays a divergence of the density of Liouvillian eigenvalues, which we call a Liouvillian Spectral Phase Transition (LSPT) drawing an analogy with an Excited-State Quantum Phase Transition (ESQPT) \cite{Caprio2008, Stransky2016, Cejnar2021} occurring in some closed quantum systems. The main dynamical consequence of an ESP is the slowing down of the relaxation to the SS. We also show that the emergence of this phase is closely linked to a dissipative quantum phase transition \cite{Biella2018}, leading to the formation of a boundary time crystal (BTC) at the critical point \cite{Iemini2018, Buca2019, picci2021, Carollo2021}.

{\em Model and spectral properties.-}  We consider the dynamics of a large collective spin $\hat{\mathbf{J}}$ \cite{Chase2008, Nori2018, Ferreira2019}, representing a set of $2j$ $1/2$-spins, subject to a uniform magnetic field and in contact with a spin-polarized bath in the GKSL approximation. This system is a particular limit of the Richardson-Gaudin model and has been solved exactly in \cite{Ribeiro2019,Lerma2020}. It is relevant for the study of quantum cavity electrodynamics \cite{Fitzpatrick2017,Shankar2021}, magnetic grains on a metallic surface \cite{Ferreira2019}, or superconducting quantum circuits \cite{Hauss2008}. The Hamiltonian of the closed system reads ($\hbar=1$)
\begin{equation}
  \hat{H} = -h \hat{J}_{z},
  \label{eq:ham}
\end{equation}
where $h$ is a local magnetic field. The Lindblad jump operators describing the coupling to the environment are
\begin{equation}
  \hat{L}_{0} = \sqrt{\frac{\Gamma_{0}}{j}}\, \hat{J}_{z},\quad
  \hat{L}_{\pm} = \sqrt{\frac{\Gamma}{j} \frac{1\mp p}{2}}\, \hat{J}_{\pm},
  \label{eq:jumps}
\end{equation}
where $\Gamma$ and $\Gamma_{0}$ define the dissipation strength, $j$ is the magnitude of the collective spin $\mathbf{\hat{J}}$ (see Ref. \cite{Ribeiro2019}) and $-1\leq p\leq 1$ can be understood as a control parameter accounting for the degree of polarization in the bath, which drives the system towards the Hamiltonian ground-state for $p < 0$ and excites it for $p > 0$. Arbitrary units are used throughout.

We express the Liouvillian superoperator in the vectorial representation \cite{Schmutz1978,Prosen2008,yoshioka2019,Cattaneo2020} where the density matrix $\rho_{\alpha,\beta}$ of dimension $\mathcal{N}\times \mathcal{N}$, with $\mathcal{N}=2j+1$, is mapped to a vector $\ket{\alpha,\beta}$ in a Hilbert space of dimension $\mathcal{N}^2$. Accordingly, the angular momentum operators acting on the left or right of the density matrix are mapped as $\mathbf{\hat{J}}\,\rho\rightarrow \hat{J}\otimes I\,|\rho \rangle =\mathbf{\hat{K}}_1|\rho \rangle$, and $\quad\rho\,\mathbf{\hat{J}}\rightarrow I\otimes \hat{J}^{T}|\rho \rangle= \mathbf{\hat{K}}_2|\rho \rangle.$

The $\mathbf{\hat{K}}_i$ operators are spin operators of equal magnitude $j$. The Liouvillian in the vectorized space becomes
\begin{equation}
  \begin{split}
  \mathcal{L} =&
  -\Gamma (j+1)
  +ih \left( \hat{K}_{1z}-\hat{K}_{2z} \right)
   + \frac{\Gamma}{j} \hat{K}_{1z} \hat{K}_{2z} \\
  & + \frac{\Gamma - \Gamma_{0}}{2j} \left( \hat{K}_{1z}-\hat{K}_{2z} \right)^{2}
  -\frac{\Gamma}{j} \frac{p}{2} \left( \hat{K}_{1z} + \hat{K}_{2z} \right) \\
  &+ \frac{\Gamma}{j} \frac{1-p}{2} \hat{K}_{1+} \hat{K}_{2+} + \frac{\Gamma}{j} \frac{1+p}{2} \hat{K}_{1-} \hat{K}_{2-}~.
  \end{split}
  \label{eq:liouv}
\end{equation}
The $z$-component of the angular momentum $\hat{K}_{z} = \hat{K}_{1z} - \hat{K}_{2z}$ is a weak symmetry \cite{vanCaspel2018} which classifies the Liouvillian eigenvalues into symmetry sectors with quantum numbers $-2j \le M \le 2j$. The term with $\Gamma_0$ represents a constant shift $\frac{-\Gamma_0}{2 j} M^2$ in the real part of the Liouvillian eigenvalues in each $M$ sector. For simplicity, we will set $\Gamma_{0}=0$ hereinafter. We denote the ordered Liouvillian eigenvalues as $\lambda_{N,M}$, with $0\leq N\leq 2j-|M|$, such that $\textrm{Re}\left( \lambda_{N,M} \right) \geq \textrm{Re}\left( \lambda_{N+1,M} \right)$. The eigenvalue of the unique SS of this model is $\lambda_{0,0}=0$ for any finite $j$.

\begin{center}
  \begin{figure}
  \includegraphics[width=\linewidth]{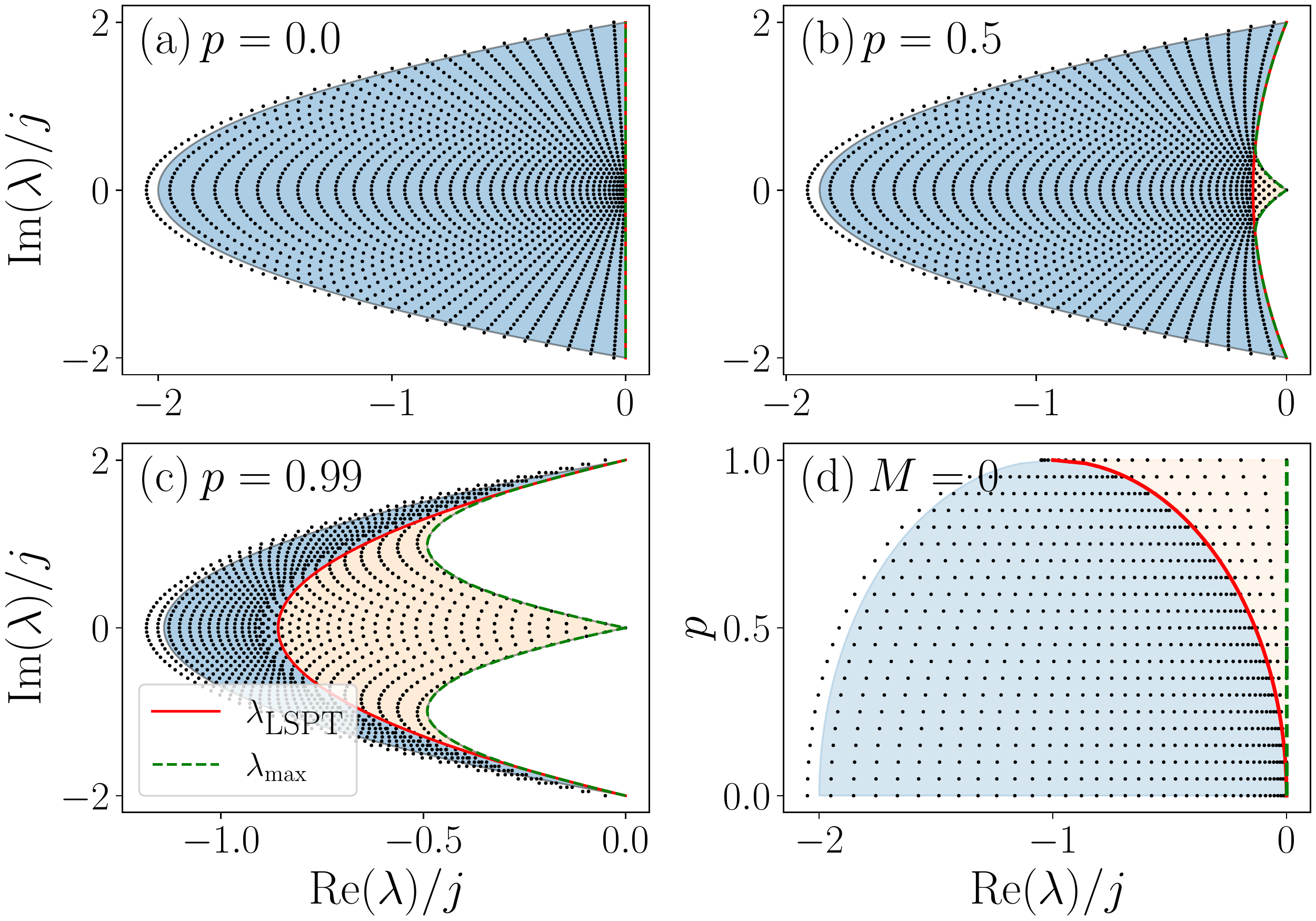}
  \caption{(a-c) Liouvillian spectrum for several values of the polarization $p=0,0.5,0.99$ for a finite spin $j=20$ computed with exact diagonalization (dots). The spectrum in the TL is divided into a region with second order EPs (orange [light] shaded region, right) and one with no spectral degeneracies (blue [dark] shaded region, left). These two regions are separated by a LSPT (solid red line). We show the eigenvalues with maximum real parts (dashed green lines) for each symmetry sector $M$ in the TL. (d) Real part of the $M=0$ spectral sector for different values of $p$.}
  \label{fig:phases}
\end{figure}
\end{center}

Fig.~\ref{fig:phases} shows the spectrum for a finite spin. The case with $p=0$, displayed in Fig.~\ref{fig:phases}(a), shows a normal spectrum without degeneracies. By contrast, the $p=0.5$ case in Fig.~\ref{fig:phases}(b) has remarkable features. The eigenvalues within a region close to the SS are quasi-degenerate in pairs (orange, light)  \cite{Ribeiro2019}, whereas no degeneracies are found in the rest of the spectrum (blue, dark)  \cite{Ribeiro2019}. The boundary between these two regions displays a close packing of eigenvalues in the region where the density of eigenvalues diverges in the TL (red line) \cite{Ribeiro2019}, defining an LSPT. The case with $p=0.99$ is displayed in Fig.~\ref{fig:phases}(c) with a growing fraction of the eigenvalues that are quasi-degenerate in pairs. In all cases, the dashed green lines represent the TL of the eigenvalues with the highest real part for each $M$ block, obtained from the semiclassical approximation in the TL of \cite{Ribeiro2019}. Finally, Fig.~\ref{fig:phases}(d) shows how the size of the degenerate spectral region changes with $p$ in the sector $M=0$, suggesting that the ESP disappears at $p=0$, while it covers the entire spectrum with the exception of the SS for $p=1$.

{\em Exceptional spectral phase.-} Quantum Liouvillian superoperators can display spectral degeneracies whereby two or more eigenvectors coalesce giving rise to EPs. To see if this occurs in the orange [light] region of Fig.~\ref{fig:phases}, we numerically study the $M=0$ sector (analogous results are obtained for other $M$).

\begin{center}
  \begin{figure}
  \hspace*{-0.4cm}\includegraphics[width=0.51\textwidth]{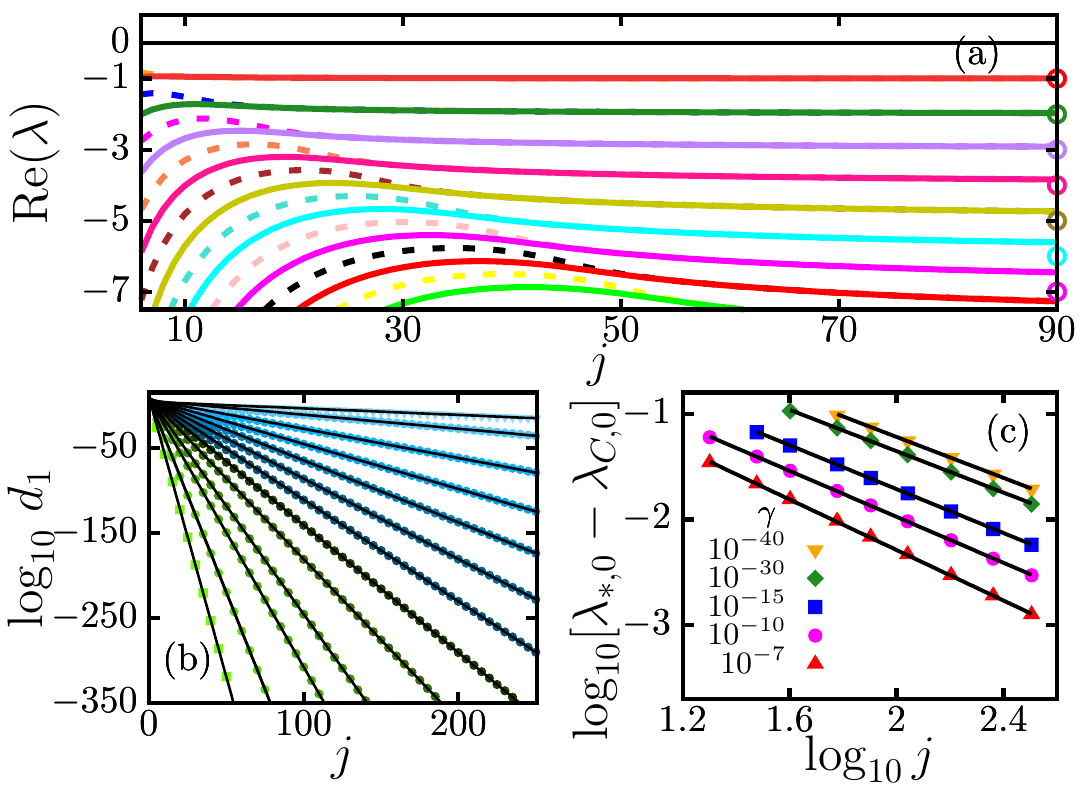}
  \caption{(a) Real part of the eighteen eigenvalues closest to the SS (0 eigenvalue) in the sector $M=0$ for $p=0.5$ as a function of the system size $j$. When $j\gg 1$ the eigenvalues agree well with the HP approximation (circles). (b) Eigenvector distance $d_{1}$ of the eigenvalue doublet closest to the SS as a function of $j$, from $p=0.999$ (green squares) to $p=0.05$ (blue triangles). (c) Scaling of the LSPT eigenvalue precursor with respect to the TL  value $\lambda_{C,0}/j= -0.133975$ for $p=0.5$ obtained from the distance between eigenvectors. Shapes represent different bounds $\gamma$. In (b), (c) lines represent the best linear fit. Parameters are $\Gamma=h=1$.}
  \label{panelespectro}
  \end{figure}
\end{center}

First, we study how the distance between neighboring eigenvalues decreases with system size.  Fig.~\ref{panelespectro}(a) shows the evolution of the eigenvalues closest to the SS as $j$ is increased, for $p=0.5$. Even for moderate values of $j$, the spectrum displays degenerate pairs. Furthermore, each doublet is well approximated by our Holstein-Primakoff (HP) expansion to first order in $j$ (TL) [see below], yielding $\lambda_{2n-1,0} = \lambda_{2n,0}= -2 |p| \Gamma n$, with $n\in\mathbb{N}$, and $\lambda_{0,0}=0$. Deviations from the harmonic HP spectrum are due to finite size effects. In Fig.~\ref{panelespectro}(b) we show how the eigenvector distance $d_{1}$ between the pair of eigenvalues closest to the SS decreases with $j$ for different values of $p$. The eigenvector distance $d_N$ between a pair of eigenvalues is defined as $d_{N}=1-||\left<N+1,0 | N,0 \right> ||$, where $\left<N+1,0 | N,0 \right>$ denotes the scalar product of the right eigenvectors. For this calculation all eigenvectors are normalized by the Euclidean norm of their elements. In all cases studied for $0<p<1$, $d_N$ decreases exponentially, producing EPs in the TL.

Because results in Fig.~\ref{panelespectro}(b) only account for a pair of degenerate eigenvalues, we come back to the case with $p=0.5$ to determine the boundary of the ESP. Following \cite{Corps2021}, we perform the following finite-size scaling: (i) we select a given bound $\gamma$ for the distance $d_{N}$, such that if $d_{N}<\gamma$ we consider that the two eigenvectors  $|N,M \rangle$ and $| N+1,M \rangle$ have coalesced; (ii) we identify a precursor of the critical eigenvalue, $\lambda_{*,M} (\gamma,j)$, as the eigenvalue $\lambda_{N+1,M}$ with largest real part in the symmetry sector $M$ fulfilling $d_{N} > \gamma$, and (iii) we study how this precursor changes with system size. Hence, in Fig.~\ref{panelespectro}(c) we display $\lambda_{*,0}(\gamma,j) - \lambda_{C,0}$, being $\lambda_{C,0}$ the TL {\em critical} value of the LSPT in the sector $M=0$ \cite{Ribeiro2019}, as a function of $j$ and for different bounds, $\gamma$. Our results show that $\lim_{j \rightarrow \infty}\lambda_{*,0}(\gamma,j)= \lambda_{C,0}$ following a power law $\lambda_{*,0}(\gamma,j)-\lambda_{C,0}\sim j^{z}$ with $z\approx -1$ for all bounds $\gamma$. A similar scaling holds for any other value of $0 < p < 1$ in the other $M$ sectors (not shown). Therefore,  {\em all the eigenvectors with eigenvalues fulfilling $\textrm{Re}(\lambda_{N,M})>Re(\lambda_{C,M})$, except for the one with largest real part, $\lambda_{0,M}$, coalesce in pairs in the TL}. This shows  that the Liouvillian spectrum in the TL is split into two different regions, whose boundary is the LSPT. We define a {\em Liouvillian spectral phase} as a region in the Liouvillian spectrum displaying some particular properties different from the rest of the spectrum, and bounded  by a non-analyticity in the density of eigenvalues. In our case, the spectrum of the Liouvillian Eq. \eqref{eq:liouv} is split into two spectral phases: a normal phase, with no degeneracies, and an {\em Exceptional Spectral Phase} of second order EPs. The size of the ESP grows with $p$ and covers the entire spectrum but the SS at $p=1$, as we show below.

{\em Exact ESP in the limit $p=\pm 1$.-} For $p=\pm 1$ the dissipative terms are only spin lowering (or raising) Lindblad jumps and the Liouvillian matrix becomes triangular. Therefore, for any system size, the Liouvillian eigenvalues of the symmetry sector $M$ are simply its diagonal elements,
\begin{equation}
  \begin{split}
    &\lambda_{m,M} =\ \langle m, m-M|\mathcal{L}| m,m-M \rangle \\
    =& -\Gamma (j+1) + ih M
    + \frac{\Gamma}{2j} \left[ M(M+p) - 2m(M-m+p) \right].
  \end{split}
\end{equation}
These eigenvalues are all exactly degenerate in pairs $\lambda_{m,M} \equiv \lambda_{M-m+p,M}$ except for: (a) $m = -j+M$ if $M\geq 0$ and $m=-j$ if $M \leq 0$, corresponding to the eigenstates with highest real part of that symmetry sector (and the SS for $M=0$); and (b) $m = \frac{M+p}{2}$ if $M$ is odd, corresponding to the eigenvalue with lowest real part. Thus, the spectrum in the limit $p=\pm 1$ for any $j$ is fully degenerate in pairs except for the extremal eigenvalues.

The action of the Liouvillian on a general eigenstate
\begin{equation}
| N,M \rangle = \sum_{m=\max \left\{ -j,-j+M \right\}}^{\min \left\{ j,j+M \right\}} \rho_{m}^{N,M} | m,m-M \rangle.
\end{equation}
with eigenvalue $\lambda_{N,M}$ is given by the eigenvalue equation
\begin{multline}
\left[ \left( \mathcal{L} - \lambda_{N,M} \right) | N,M \rangle \right]_{m} =\\
\left( \lambda_{m,M} - \lambda_{N,M} \right) \rho_{m}^{N,M} + c_{m+1}\rho_{m+1}^{N,M} = 0,
\end{multline}
with $c_{m} = \langle m-1, m-1-M|\mathcal{L}| m,m-M \rangle$. This eigenvalue equation \textit{admits only one solution}, which for $p=1$ is
\begin{equation}
\rho_{m}^{N,M} =
\begin{cases}
\prod_{i=m}^{N-1}\frac{c_{i+1}}{\lambda_{N,M}-\lambda_{i,M}} & \quad m < N \\
1 & \quad m = N \\
0 & \quad m > N
\end{cases},
\end{equation}
while a similar solution holds for $p=-1$. Thus, for every pair of degenerate eigenvalues, $\lambda_{N,M}\equiv\lambda_{1+M-N,M}$, the dimension of the kernel of $\mathcal{L} - \lambda_{N,M}\mathbf{I}$, with $\mathbf{I}$ the identity matrix, is $1$. Therefore, the Liouvillian at the limits $p=\pm 1$ is non diagonalizable, almost every eigenvalue is exactly degenerate in pairs and every pair forms a second order EP.

\textit{Dynamical slowing-down in the exceptional phase.-}  As a consequence of the appearance of EPs, the number of eigenvectors of the Liouvillian is smaller than the dimension of the Hilbert space, ${\mathcal N}^2$. Hence, generalized eigenvectors of rank $2$ $\left\vert \overline{N,M}\right\rangle $ are required to span the complete Hilbert space. Their dynamical relevance in an ESP can be measured by the dimension of the space $\mathcal{D}$ spanned by such generalized eigenvectors. From our previous results, we infer that $0 < \mathcal{D} /{\mathcal N}^2 < 1/2$ for $0<p<1$ in the TL; the larger $p$, the larger the ratio $\mathcal{D}/{\mathcal N}^2$. We therefore expect that the time evolution of a large variety of initial conditions will be influenced by these generalized eigenvectors.

To study the consequences of the EPs precursors in finite-size systems we employ the HP expansion in the TL defining an initial density matrix composed by the SS and its closest doublet in the $M=0$ sector. To compute these slow decaying states we map the collective spin operators to a bosonic space for e.g. $p > 0$, $\hat{K}_{i+}= b_{i}^{\dagger}\,\sqrt{2j-b_{i}^{\dagger}b_{i}} = \left( \hat{K}_{i-} \right)^\dagger$ and $\hat{K}_{iz}=-j+b_{i}^{\dagger}b_{i}$, with $b_i,\,i\in\{1,2\}$ standard boson annihilation operators, and keep the highest-order terms in $j$. A non-unitary ($\mathcal{L}$ is non-hermitian) Bogoliubov transformation $\beta_{1}=u\hat{b}_{1}-v\hat{b}_{2}^{\dagger},\,\beta_{2}=u\hat{b}_{2}-v\hat{b}_{1}^{\dagger}$ [$\overline{\beta}_{1}=\overline{u}\hat{b}_{1}^{\dagger}-\overline{v}\hat{b}_{2},\,\overline{\beta}_{2}=\overline{u}\hat{b}_{2}^{\dagger}-\overline{v}\hat{b}_{1}$] with $u=(1+p)/(2\sqrt{2})$, $v=(1-p)/(2\sqrt{p})$, and $\overline{u}=\overline{v}=1/\sqrt{p}$, such that $[\beta_{i},\overline{\beta}_{j}]=\delta_{ij}$, diagonalizes the highest-order Liouvillian, $\mathcal{L}_{h}/\Gamma=(ih/\Gamma-p)\overline{\beta}_{1}\beta_{1}-(ih/\Gamma+p)\overline{\beta}_{2}\beta_{2}$. Therefore, the SS is the vacuum state of the quasiboson operators satisfying $\mathcal{L}_{h}\ket{0,0}=0$,
\begin{equation}
    | 0,0 \rangle = \sum_{n=0}^{2j} \frac{\alpha^{n}}{n!} \left( b_{1}^\dagger b_{2}^\dagger \right)^{n}  | 0 \rangle = \frac{1}{\mathcal{Z}}\sum_{m=-j}^{j}\alpha^{m+j}\ket{m,m},
    \label{eq:SSp0}
\end{equation}
with $\alpha = \frac{1-p}{1+p}$ and $\mathcal{Z}$ a normalization constant. Every other eigenstate is generated by successive application of the quasiboson creation operators onto the SS, while its corresponding generalized eigenvector satisfies $\left( \mathcal{L}_{h} - \lambda_{N,M} \right)| \overline{N,M} \rangle = | N,M \rangle$. For example, in the symmetry sector $M=0$, the slowest decaying eigenvector is $\ket{1,0}=4p/(1+p)^{2}\hat{b}_{1}^{\dagger}\hat{b}_{2}^{\dagger}\ket{0}-2/(1+p)\ket{0}$, and its generalized eigenvector, $\ket{\overline{1,0}}=\sum_{n=0}^{2j}\Omega_{n}/n! (\hat{b}_{1}^{\dagger}\hat{b}_{2}^{\dagger})^{n}\ket{0}$ with $\Omega_{0}=1,\,\Omega_{1}=-(1+2p)/(1+p),\,\Omega_{2}=p/(1+p)^{2},$ and $\Omega_{n}=\Omega_{2}[2p/(1+p)]^{n-2}\binom{n}{2}^{-1},\,n\geq 3$.

Starting from an initial density matrix $\left|\rho(t=0)\right> = \left| 0,0 \right> + a \left| 1,0 \right> + b \left| \overline{1,0} \right>$, with $a,b$ ensuring physicality of the density matrix, the time evolution in the TL is given by \cite{Weintraub2008}
\begin{equation}
  \left|\rho(t)\right>=\left| 0,0 \right> + \left( a + b t \right) e^{\lambda_{1,0} t} \left| 1,0 \right>  + b e^{\lambda_{1,0} t} \left| \overline{1,0} \right>,
  \label{eq:partner}
\end{equation}
where $\lambda_{1,0}=-2|p|\Gamma$ is the dissipative gap in the TL.  Consequently, the relaxation to the SS is slowed down by a linear correction, proportional to the overlap between the initial condition and $\left|\overline{1,0} \right >$. This correction is quite significant at short times. For example, for a decay constant of $\lambda_{1,0}=-2$, the time for the population of the eigenstate $| 1,0 \rangle$ to be reduced by half would be $t_{1/2}=0.3466$ for normal eigenstates, while for an EP it would be $t_{1/2}=0.5731$, implying a $65\%$ increase, if both the eigenstate and its corresponding generalized eigenvector of rank $2$ are populated.

\begin{figure}[tp]
  \centering
  \includegraphics[width=0.95\linewidth]{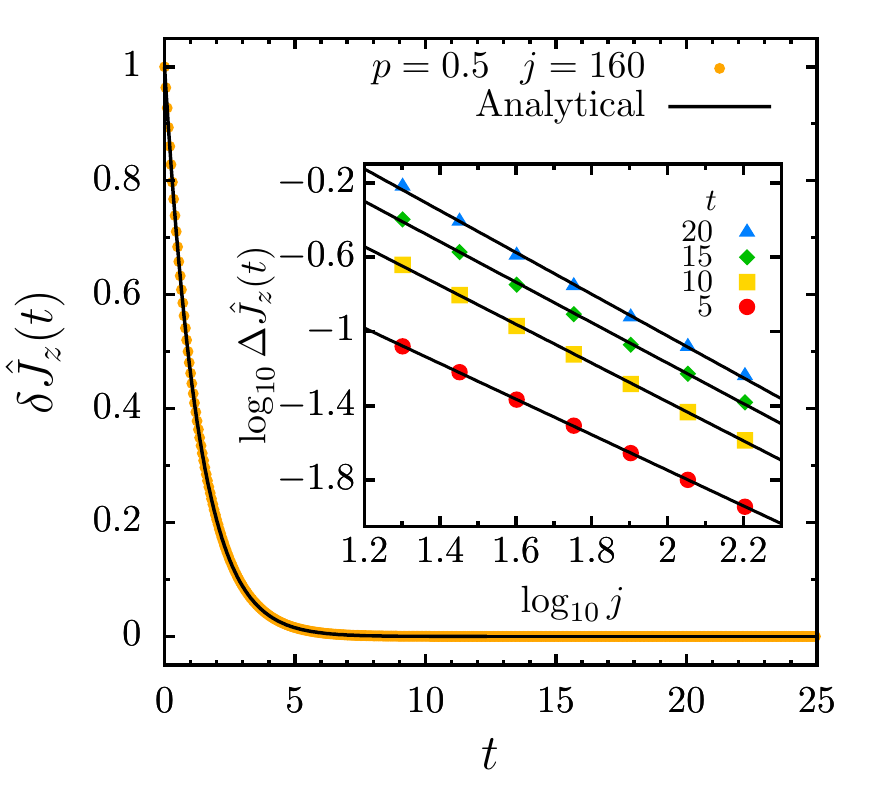}
  \caption{Time evolution of $\delta \hat{J}_{z}(t)$ from the initial condition $\ket{\rho(0)}$ \eqref{eq:partner} with $a=0$ and $b=1/6$ for $p=0.5$ and $j=160$. Circles represent the purely exponential evolution for $j=160$ and the solid line is the theoretical curve [Eq. \eqref{eq:partner}]. Inset: comparison of the difference between numerics and theory, $\Delta \hat{J}_{z}(t)$, for different system sizes. Lines represent the best linear fit to the points, revealing a power-law behavior $\Delta \hat{J}_{z}(t)\sim j^{-a}$, $a=O(1)$. }
  \label{figssjp}
\end{figure}

In Fig.~\ref{figssjp} we numerically compute the time evolution of $\hat{J_z}$ starting from this initial state {\em for a   finite-size system with $j=160$}, and compare with the expected value in the TL given by Eq. \eqref{eq:partner}. In the main panel, we represent
 $\delta \hat{J}_{z}(t)=\frac{\langle\hat{J}_{z}(t)\rangle-\langle \hat{J}_{z}(\infty)\rangle}{\langle \hat{J}_{z}(0)\rangle - \langle \hat{J}_{z}(\infty)\rangle},$
where we calculate the time evolution of finite size systems by numerically computing the exponential matrix $e^{\mathcal{L}\cdot t}$. The time evolution of $\hat{J}_z(t)$ is almost indistinguishable from the analytical curve in the TL at any time. Even for finite $j$, the slowing-down in the relaxation to the SS is comparable to that of a true EP. This is reflected in the inset of Fig.~\ref{figssjp} by  the relative difference between the finite size and theoretical time evolution curves (computed in the TL), $\Delta \hat{J}_{z}(t)=(\delta \hat{J}_{z}(t)-\delta \hat{J}_{z}(t)_{\textrm{theo}})/\delta \hat{J}_{z}(t)_{\textrm{theo}}$, as a function of $j$ and for different fixed times $t$. The black lines are a fit of the numerical points to a power-law behavior $\Delta \hat{J}_{z}(t)\sim j^{-a}$ with $a=O(1)$. As we observe in the inset, the relative difference between the evolution for finite $j$ and that of an EP becomes larger as time increases but is rapidly reduced for bigger system sizes, as the distance between neighboring eigenvalues drops.

Summarizing, the presence of an ESP with a large fraction of quasi-EPs slows down the evolution to the SS for a large variety of initial conditions having a significant overlap with the set of generalized eigenvectors.

{\em Steady state and dissipative quantum phase transition.-} We now focus on the SS, Eq. \ref{eq:SSp0}, in the TL. Writing this state in the original space of density matrices, it coincides with the canonical equilibrium ensemble for the Hamiltonian Eq. \eqref{eq:ham}, $\rho_{SS} \equiv \exp(-\beta H)/Z$, with an inverse temperature $\beta = \frac{1}{h} \ln \left( \frac{1-p}{1+p}\right)$ and a partition function $Z=e^{-\beta h j}(1-e^{\beta h (2j+1)})/(1-e^{\beta h})$, enabling an easy computation of all thermodynamic properties of the SS. In particular, the magnetization for an arbitrary polarization of the bath in the TL can be evaluated as $\langle \hat{J}_z\rangle_{SS}=(1/h) \partial \ln Z/\partial \beta$,
\begin{equation}
  \lim_{j \rightarrow \infty} \frac{\langle \hat{J}_z \rangle_{\textrm{SS}}}{j} = \begin{cases} -\textrm{sign} \left( p \right), \; &p \neq 0, \\ 0, \; &p =0. \end{cases}
\end{equation}

The abrupt change in the sign of the magnetization implies the existence of a dissipative QPT (DQPT) at $p=0$ already discussed in \cite{Ribeiro2019}. At the critical point $p=0$, the Liouvillian Eq. \eqref{eq:liouv} conserves the total angular momentum and therefore, in addition to the $z-$component $\hat{K}_{z}$, it commutes with  $\mathbf{\hat{K}}^2 = \left(\mathbf{\hat{K}}_{1} - \mathbf{\hat{K}}_{2}\right)^2$,  closing a O(3) algebra, and allowing for a complete analytical solution \cite{Claeys2022} of the Liouvillian spectrum
\begin{equation}
  \lambda_{K,M} = i h M + \frac{\Gamma}{2j} M^2 - \frac{\Gamma}{2j} K(K+1).
  \label{eq:rotor}
\end{equation}
Here, $K$ and $M$ are quantum numbers given by $\mathbf{\hat{K}}$ and $\hat{K}_{z}$, respectively, satisfying $0\leq K\leq 2j$ and $-K\leq M\leq K$. For each $M$, $K=|M|$ gives the eigenvalue with largest real part. Note that $\lambda_{K,M=0}\in\mathbb{R}$ while $\lambda_{K,M\neq 0}\in\mathbb{C}$ with $\textrm{Im}(\lambda_{K,M})=hM$. The completeness of the exact solution precludes the existence of EPs. When the imaginary term $ihM$ is discarded, Eq. (\refeq{eq:rotor}) represents the spectrum of an axially symmetric rotor with $j$ acting as the moment of inertia. In the TL such moment of inertia diverges and the O(3) symmetry may be broken, implying that: (i) {\em an infinite number of Liouvillian eigenstates, $\ket{K,0}$, become degenerate with the SS}, and (ii) {\em analogously for $M \neq 0$, an infinite number of eigenstates $\ket{K,M}$ have 0 real part but oscillate with a period $T=2 \pi/(h M)$}. These two properties characterize the critical phase as a BTC. Moreover, from the complete set of Liouvillian eigenstates $\ket{K,M}$ in the coupled basis, we obtain the eigenmatrices by decoupling with a Clebsch-Gordan coefficient
\begin{equation}\begin{split}
\ket{m}\bra{m'}&\equiv (-1)^{j-m'}\ket{m,-m'}\\&=\sum_{K,M}(-1)^{j-m'}\bra{jm,j-m'}\ket{K,M}\ket{K,M},
\end{split}
\end{equation}
with $M=m-m^{\prime }$. The density matrix can be evolved in time using the exact spectrum Eq. \eqref{eq:rotor} and  $\rho(t)=e^{\mathcal{L} t} \rho(0)$. For an initial density matrix $\rho(0)=\sum_{mm'}\rho_{mm'}\ket{m}\bra{m'}$,

\begin{equation}\begin{split}
&\rho(t)=\sum_{mm',KM,nn'}\rho_{mm'}e^{[ihM+\frac{\Gamma}{2j}M^{2}-\frac{\Gamma}{2j}K(K+1)]t}(-1)^{j-m'}\\&\times\bra{jm,j-m'}\ket{K,M}\bra{jn,j-n'}\ket{K,M}(-1)^{j-n'}\ket{n}\bra{n'}.
\end{split}
\end{equation}
For an observable $\hat{O}$, the expectation value $\langle \hat{O}(t)\rangle=\Tr[\rho(t)\hat{O}]$ equals

\begin{equation}\begin{split}
&\langle \hat{O}(t)\rangle=\sum_{mm',KM,nn'}\rho_{mm'}\bra{n'}\hat{O}\ket{n}e^{[ihM +\frac{\Gamma}{2j}M^{2}-\frac{\Gamma}{2j}K(K+1)]t}\\&\times(-1)^{j-m'}\bra{jm,j-m'}\ket{K,M}(-1)^{j-n'}\bra{jn,j-n'}\ket{K,M}.
\end{split}
\end{equation}

\begin{center}
  \begin{figure}
  \hspace*{-0.85cm}\includegraphics[width=1.2\linewidth]{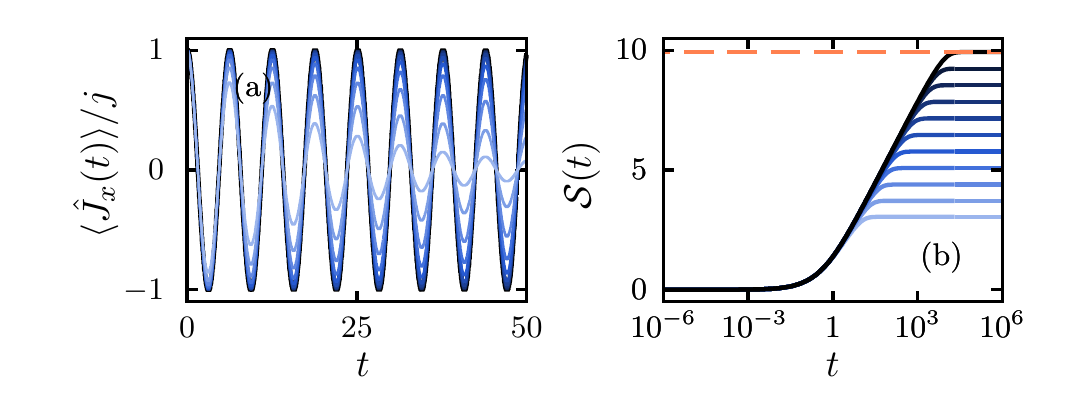}
  \caption{Dynamics at $p=0$ for the system with $\Gamma=h=1$. (a) Time evolution of  $\langle \hat{J}_x \rangle$ for an initial state with maximum $\langle \hat{J}_x(t=0)\rangle$, computed from Eq. \eqref{eq:jx}. From light to dark, $j=10\times 2^{k},\, k=0,\dots,10$. (b) Entropy evolution in time computed from for a pure initial state $\rho(0)=\ket{j}\bra{j}$; the dashed line represents the upper bound of the entropy for $j=10\times 2^{10}$.}
  \label{paneldinamica}
  \end{figure}
\end{center}

Using the expressions of the angular momentum matrix elements in terms of Clebsch-Gordan coefficients and the corresponding orthogonality properties, the result for $\left\langle \hat{J}_z(t)/j \right\rangle$ is
\begin{equation}
  \left< \hat{J}_z(t)/j \right> = \left< \hat{J}_z(0)/j\right> e^{-\frac{\Gamma}{j}t} \xrightarrow[j \rightarrow \infty]{} \left< \hat{J}_z(0)/j\right>,
  \label{eq:jz}
\end{equation}
implying that $\left< \hat{J}_z(t)/j \right>$ stays frozen in its initial value in the TL \cite{Sanchez2019}, {\em with this value depending only on the initial state}. This is a consequence of (i), implying that the non-unitary evolution conserves the initial value of $\left< \hat{J}_z(t)/j \right>$ if the bath is not polarized.

The consequences of (ii) can be explored by studying the dynamics of $\left\langle J_x/j \right\rangle$,
\begin{equation}
  \left<\hat{J}_x(t)/j\right> = \left< \hat{J}_x(0)/j \right> e^{-\frac{\Gamma}{2j} t}\cos(ht) \xrightarrow[j \rightarrow \infty]{} \left< \hat{J}_x(0)/j \right> \cos(ht).
  \label{eq:jx}
\end{equation}
This implies that $\left< \hat{J}_x(t)/j \right>$ does not reach a SS value; instead, it remains oscillating with no damping in the TL. In Fig.~\ref{paneldinamica}(a) we illustrate how the damping of $\left< \hat{J}_x(t) /j\right>$ is reduced as we increase the value of $j$ towards the TL.

The critical point $p=0$ of the dissipative QPT, having the unusual property of a BTC, is the starting point of a LSPT that extends into the region $0 < |p| < 1$ and serves as the boundary of the ESP. A similar phenomenon usually links QPTs and ESQPTs in closed systems \cite{Caprio2008,Stransky2016, Cejnar2021}.

Finally, we show how dissipation dominates the dynamics at $p=0$ despite the previous results. We display in Fig.~\ref{paneldinamica}(b) the entropy as a function of time $\mathcal{S}(t)=\textrm{Tr}\,\left[\rho(t) \ln \rho(t)\right]$ of a pure initial state $\rho(0)=\ket{j}\bra{j}$, where we obtain $\rho(t)$ by numerically computing the exponential matrix $e^{\mathcal{L}\cdot t}$. The dissipation induces the growth of the entropy at a rate depending only on the system parameters until it saturates at the maximum entropy $\mathcal{S}_{\max}=\ln \left( 2j+1 \right)$. Even when the initial value of $\left\langle \hat{J}_z \right\rangle$ is conserved, the entropy grows boundlessly in the TL.

\textit{Conclusions.-} In this Letter, we propose the existence of an Exceptional Spectral Phase in a system composed by a collective spin interacting  with a polarized bath and subject to a magnetic field. The main consequence is that the relaxation towards the SS is slowed down in the thermodynamic limit when the initial state is spread over this phase. The exceptional  phase is separated from the rest of the Liouvillian spectrum by a critical line in which the density of eigenvalues diverges, which constitutes the generalization of an excited-state quantum phase transitions to open quantum systems. We also show that this critical line is transferred onto the SS when the bath is not polarized, giving rise to a dissipative quantum phase transition and to the formation of a BTC. Due to the widespread presence of ESQPTs in collective spin models, it is reasonable to expect that these unique features should not be exclusive to this particular model. We believe that it is then of great interest to investigate more general scenarios of dissipative collective spin systems searching for similar properties \cite{picci2021}.

\begin{acknowledgments}
This work was partially supported by grants PGC2018-094180-B-I00 and PID2019-104002GB-C21 funded by MCIN/AEI/10.13039/501100011033 and FEDER "A way of making Europe". It has been also supported by the CAM/FEDER Project No. S2018/TCS-4342 (QUITEMAD-CM) and by the Consejer\'{i}a de Econom\'{i}a, Conocimiento, Empresas y Universidad de la Junta de Andaluc\'{i}a under projects UHU-1262561 and P20-00764. This research has also been supported by CSIC Research Platform on Quantum Technologies PTI-001. A. L. C. acknowledges financial support from `la Caixa' Foundation (ID 100010434) through the fellowship LCF/BQ/DR21/11880024.

\textit{Author contributions}: A. R.-G. and J.D. conceived the project. A.R.-G., J.D. and A.L.C. performed the analytical calculations. A.L.C., A.R.-G., A.R. and R.A.M. performed the numerical calculations, with the collaboration of F. P.-B. and J. E. G.-R. All figures were formated by A. L. C. and A.R.-G. Finally, R.A.M., A.R. and A.L.C. took the lead in writing the manuscript with A.R.-G. and J.D. contributing significantly to the final version. All authors contributed to the discussions and have agreed on the final version of the manuscript.
\end{acknowledgments}

\end{document}